\documentclass[12pt]{article}
\pdfoutput=1
\usepackage{graphicx}
\usepackage{color}
\usepackage{amsmath}
\usepackage{multirow}
\usepackage[utf8]{inputenc}

\usepackage{mathptmx}

\newenvironment{sciabstract}{%
\begin{quote} \bf}
{\end{quote}}

\title{The fate of carbon dioxide in water-rich fluids at extreme conditions}
\author{Ding Pan$^{1\ast}$, Giulia Galli$^{1, 2}$\\
\normalsize{$^1$The Institute for Molecular Engineering, the University of Chicago, Chicago, IL 60637, USA}\\
\normalsize{$^2$Argonne National Laboratory, Argonne, IL 60439, USA}\\
\normalsize{$^\ast$ dingpan@uchicago.edu}
}

\date{}

\begin{document}

\maketitle

\begin{sciabstract}
Investigating the fate of dissolved carbon dioxide under extreme conditions is critical to understanding the deep carbon cycle in the Earth, 
a process that ultimately influences global climate change. 
We used first-principles molecular dynamics simulations to study carbonates 
and carbon dioxide dissolved in water at pressures (P) and temperatures (T) 
approximating the conditions of the Earth's upper mantle. 
Contrary to popular geochemical models assuming that 
molecular CO$_2$(aq) is the major carbon species present in water under deep earth conditions, 
we found that at 11 GPa and 1000 K carbon exists almost entirely 
in the forms of solvated carbonate (CO$_3^{2-}$) and bicarbonate (HCO$_3^-$) ions, 
and that even carbonic acid (H$_2$CO$_3$(aq)) is more abundant than CO$_2$(aq). 
Furthermore, our simulations revealed that ion pairing between Na$^+$ and 
CO$_3^{2-}$/HCO$_3^-$ is greatly affected by P-T conditions, 
decreasing with increasing pressure at 800$\sim$1000 K. 
Our results suggest that in the Earth's upper mantle, 
water-rich geo-fluids transport a majority of carbon 
in the form of rapidly interconverting CO$_3^{2-}$ and HCO$_3^-$ ions, 
not solvated CO$_2$(aq) molecules.

\end{sciabstract}

\section*{Introduction}
The threat of global warming and climate change makes the understanding of the Earth carbon cycle a critical and pressing task. 
In the Earth's lithosphere, sedimentary carbonates are considered as the largest carbon hosts \cite{falkowski2000global}, 
 with
carbon transport between the Earth's surface and interior \cite{manning2014geochemistry} occurring continuously  over millions to billions years \cite{depaolo2015sustainable}. 
For example, carbon belonging to organic matter and carbonate minerals may go deep into the Earth's mantle within subduction zones,
and it may be brought back to the Earth's surface by volcanism through CO$_2$ degassing \cite{dasgupta2010deep}. 
The deep carbon cycle substantially influences the carbon 
budget near the Earth's surface, which in turn impacts our energy consumption and global climate change \cite{dasgupta2010deep}.
However, whether carbon accumulates in our planet's interior is still  the subject of debate, and 
the chemical reactions leading to the predominant species of carbon
in the deep Earth are not yet well understood; hence, 
the means by which carbon transport occurs within the Earth has not yet been established \cite{manning2014geochemistry}.

\begin{figure}
\centering
\vspace{5mm}
\includegraphics[width=0.5 \textwidth]{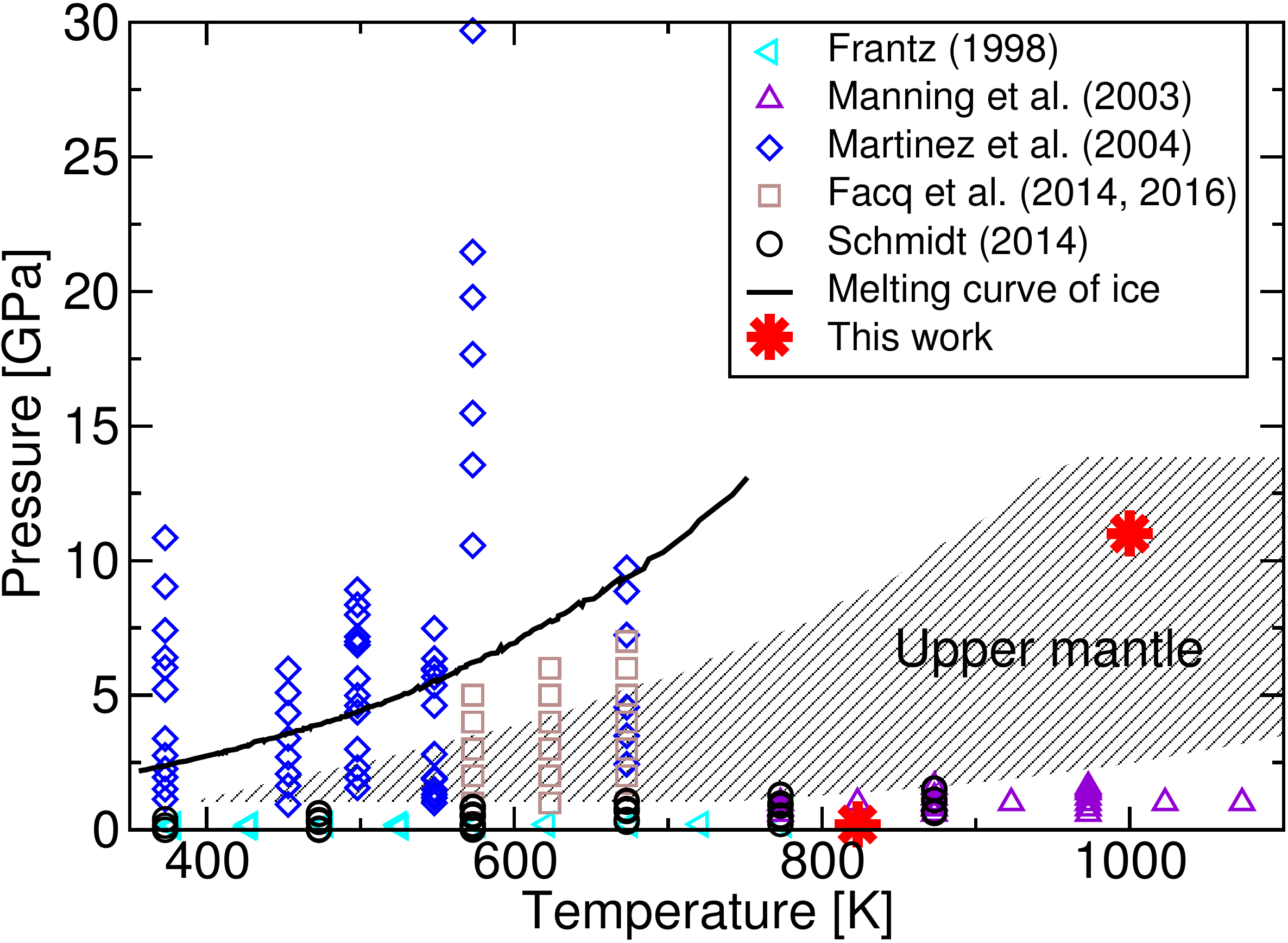}
\caption{High pressure and high temperature conditions reached in experiments on dissolved carbon in supercritical water, 
together with the conditions simulated in the present work:
Frantz (1998)\cite{frantz1998raman}, Manning et al. (2003)\cite{caciagli2003solubility}, 
Martinez et al. (2004) \cite{martinez2004high}, Facq et al. (2014, 2016) \cite{facq2014situ, facq2016carbon}, 
and Schmidt (2014) \cite{schmidt2014raman}. 
The melting curve of ice is from Ref. \cite{datchi2000extended}. 
The shaded area shows the P-T condtions of the upper mantle \cite{thompson1992water}. 
}
\label{P-T-exp}
\end{figure}

Carbon may be transported by crustal and mantle fluids in subduction zones \cite{kelemen2015reevaluating}, 
where water is a major transport medium \cite{liebscher2010aqueous}.
However the forms of dissolved carbon in water under the deep Earth conditions are poorly known. 
In the Earth's upper mantle, 
the pressure (P) may reach $\sim$ 13 GPa and the temperature (T) $\sim$ 1700 K \cite{thompson1992water}.
Experimentally it is challenging to reproduce those conditions and study aqueous solutions of CO$_2$ and carbonates in the laboratory, 
because water becomes highly corrosive \cite{weingartner2005supercritical, lin2005melting} 
and it is difficult to obtain clear vibrational spectroscopic signals, 
which are commonly used to characterize aqueous solutions at ambient conditions. 
In addition, the interpretation of existing Raman spectra measured at high P-T is controversial, as it is yet unclear how to relate
the solute concentration to the observed signals in supercritical conditions \cite{frantz1998raman,schmidt2014raman}.  
Experimental data have been obtained \emph{either} at high P (HP) \emph{or} at high T (HT), 
but it has not yet been possible to reach \emph{both} HP \emph{and} HT, as shown in Fig. \ref{P-T-exp}.  
Therefore our understanding of the chemistry of aqueous carbon is 
limited to P-T conditions of shallow mantle environments \cite{manning2013chemistry}. 
Due to the lack of experimental data, traditional geochemical models were based on several assumptions, 
one of which is that CO$_2$(aq) is the major dissolved carbon species 
in supercritical water under oxidizing conditions(e.g., \cite{duan2006equation, zhang2009model, huizenga2001thermodynamic} and references therein).

In this article, by conducting extensive first-principles molecular dynamics (MD) simulations, 
we studied Na$_2$CO$_3$ and CO$_2$ aqueous solutions at 11 GPa and 1000 K, i.e. at pressures similar to those at the bottom of the Earth’s upper mantle. 
We note that ab initio simulations, which do not require any experimental input, 
have been shown to be powerful tools to study chemical reactions under extreme conditions 
\cite{gygi2005ab}.
However they are computationally demanding and cannot yet be used to carry out a detailed study of the whole P-T range in the Earth mantle. 
The present study focuses instead on general properties of dissolved carbon in water under extreme conditions 
that could not been addressed before by simple models 
which do not take into account explicitly the molecular nature of water and the way it dissociates under pressure.
Contrary to the assumption that CO$_2$(aq) is the major carbon species under oxidizing conditions, 
we found that most of the dissolved carbon is in the form of CO$_{3}^{2-}$ and HCO$_{3}^-$, 
with continuous transformation of the two species into each other on the scale of picoseconds. 
Surprisingly, even H$_2$CO$_3$(aq) is more abundant than CO$_2$(aq) when CO$_2$ is directly dissolved into water at 11 GPa and 1000 K. 
We also found that ion pairing occurs between Na$^+$ and CO$_{3}^{2-}$ or HCO$_{3}^-$. 
Our study confirms that the molecular structure of supercritical water is markedly different from that of the liquid at ambient conditions, and shows that 
the fast chemical dynamics in aqueous solutions at extreme conditions is key to understanding dissolved carbon in deep geofluids. 
Our findings suggest that at the bottom of the upper mantle, solvated CO$_2$(aq) is hardly present in water, 
with most of the species being carbonate and bicarbonate ions.
We expect these highly active ionic species to be involved in the carbon recycle process in  subduction zones in the deep Earth, 
and to play an important role in the transport of carbon.

\section*{Results and discussion}
 
\emph{Validation of computational strategy---}Before carrying out simulations at the Earth's mantle conditions, we validated our computational strategy, 
in particular we tested the accuracy of exchange-correlation (xc) functionals used in density functional theory (DFT), 
by simulating a carbonate solution at 0.2 GPa and 823 K, 
for which experimental spectroscopic studies and  geochemical models are available \cite{johnson1992supcrt92}. 
We considered one Na$_2$CO$_3$ molecule and 62 water molecules (molality: 0.9 m) 
in a cubic simulation box with side length of 14.42 \AA. 
According to Duan et al.'s equations of state \cite{zhang2005prediction, zhang2009model} which are fitted to experimental data,
the density of water in these conditions corresponds to a pressure of $\sim$ 0.2 GPa.
We chose sodium as a representative cation since Na$_2$CO$_3$ has a large solubility in supercritical water 
\cite{martinez2004high, schmidt2014raman, pan2013dielectric}; 
Na is the seventh most abundant element in the Earth's mantle \cite{jackson2000earth}. 

\begin{figure}
\centering
\vspace{5mm}
\includegraphics[width=0.9 \textwidth]{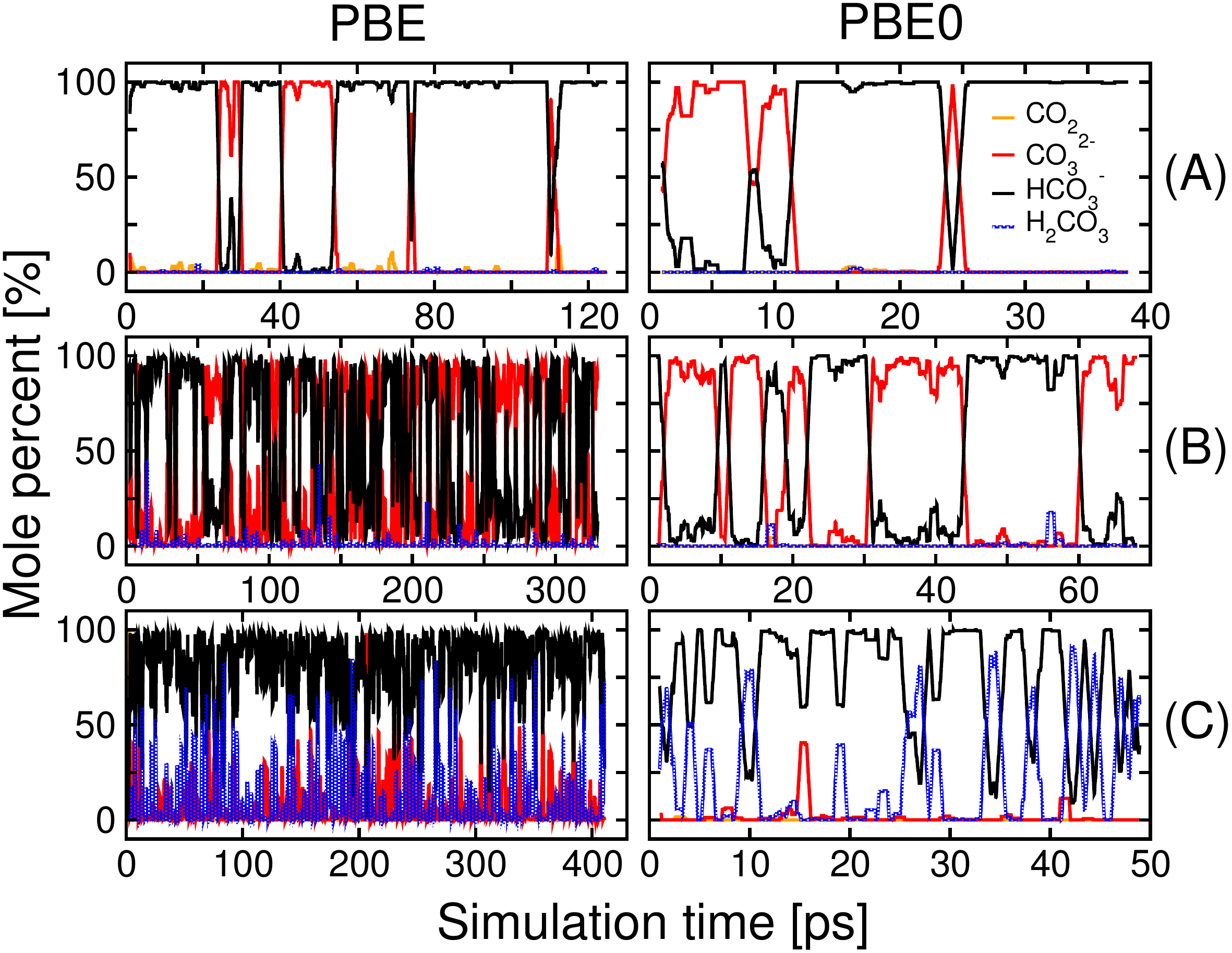}
\caption{Mole percents of CO$_2$(orange), CO$_3^{2-}$(red), HCO$_3^-$(black) and H$_2$CO$_3$(blue) as functions of simulation time 
in first-principles molecular dynamics simulations.
(A) 0.9 m (molality) Na$_2$CO$_3$ solution at 0.2 GPa and 823 K; (B) 0.9 m Na$_2$CO$_3$ solution at 11 GPa and 1000 K; (C) 0.9 m CO$_2$ solution at 11 GPa and 1000 K. 
Calculations carried out with two exchange-correlation functionals: PBE \cite{PBE} and PBE0 \cite{adamo1999toward} are compared.}
\label{mole-percent}
\end{figure}

Fig. \ref{mole-percent}(A) shows the mole percents of CO$_2$(aq), CO$_3^{2-}$, HCO$_3^{-}$ 
and H$_2$CO$_3$(aq) species as functions of simulation time 
in first-principles MD simulations carried out with semilocal (PBE \cite{PBE}) and hybrid (PBE0 \cite{adamo1999toward}) functionals.
The overall mole percents in our extensive MD simulations are summarized in Table. \ref{mol-tab}.
Using PBE, we found that bicarbonate is the dominating carbon species in the solution, with
$\sim$82 \% (mole percent of total dissolved carbon, see also methods) bicarbonate ions and $\sim$17\%  carbonate ions
 over a $\sim$130 ps MD simulation. 
The amounts of CO$_2$(aq) and H$_2$CO$_3$(aq) were negligible. 
The lifetime of bicarbonate ions exceeded 10 ps in our simulations
while the lifetime of CO$_3^{2-}$ was much shorter.

It is known that generalized gradient approximations (GGA)
may not be accurate in the description of solvated doubly charged anions, due to the charge delocalization error \cite{cohen2008insights}.
For example, in sulfuric acid aqueous solutions at ambient conditions, 
Wan et al. found that the concentration of SO$_4^{2-}$ was seriously overestimated
when using the PBE functional, while the hybrid functional PBE0, which better describes the charge localization of solvated anions \cite{gaiduk2014structural}, 
gave results in qualitative agreement with experiments \cite{wan2014electronic, guidon2008ab, distasio2014individual}.
Hence we compared PBE and PBE0 simulations to extract robust trends from our results.  
We started a PBE0 simulation from a $\sim$40 ps long PBE one. 
As shown in Fig. \ref{mole-percent}(A), HCO$_{3}^{-}$ ions transformed immediately into CO$_{3}^{2-}$ anions.
In the subsequent simulation, the total mole percent of HCO$_3^{-}$ was 27\%, indicating that PBE overestimates the concentration of  HCO$_3^{-}$ with respect to PBE0.
However, in both PBE and PBE0 simulations we found that few  CO$_2$(aq) or H$_2$CO$_3$(aq) molecules were present in the solution at 0.2 GPa and 823 K. 
Hence the absence of these species in solution was considered to be a robust result of our first-principles MD simulations.

Experimentally, Frantz studied a  one-molal K$_2$CO$_3$ solution at the same P-T conditions as those considered here, 
by Raman spectroscopy \cite{frantz1998raman}.
The concentration of a given species ($C_i$) was related to the measured Raman intensity ($I_i$) by 
$I_i = \gamma_i C_i$, where $\gamma_i$ is the Raman scattering cross section.
Based on the concentrations calculated by geochemical models,
Frantz estimated that the ratio $\frac{\gamma_{carbonate}}{\gamma_{bicarbonate}}$ decreased by $\sim$75\% 
when increasing T from 523 K to 823 K \cite{frantz1998raman}. 
Recently, after re-examining Frantz's spectroscopic data, 
Schmidt argued that such ratio should change by less than 10\% in that temperature range \cite{schmidt2014raman},
indicating that the species concentrations obtained by geochemical models in Ref. \cite{frantz1998raman} may be questionable.
Our theoretical concentrations are closer to the results obtained by the $\gamma$ ratio proposed by Schmidt \cite{schmidt2014raman}
than by Frantz \cite{frantz1998raman}.
Using Schmidt's $\gamma$ ratio and the published Raman spectra \cite{frantz1998raman}, 
we computed the mole percents of oxidized 
carbon in the solution, and found 
68\% HCO$_3^-$ and 32\% CO$_3^{2-}$ ( no CO$_2$(aq) related Raman peaks were reported in Frantz's study).
Therefore PBE appears to overestimate the amount of  HCO$_3^{-}$ by $\sim$20\% relative to the value obtained 
from Frantz's spectroscopic data\cite{frantz1998raman} and the $\gamma$ ratio proposed by Schmidt \cite{schmidt2014raman};
however the overestimate is not as serious as, e.g. for sulfuric acid aqueous solutions at ambient conditions \cite{wan2014electronic}.
This finding is consistent with our previous investigation of water under pressure, where we found that PBE 
performs better at extreme than at ambient conditions 
in terms of equation of state of water, and in predicting static and electronic dielectric constants \cite{pan2013dielectric, pan2014refractive}.
On the other hand, our PBE0 simulation yield results much closer to the experimental data. 
Our validation strategy led us to conclude that the use of the PBE functional is justified at high P and T, 
where, however, some additional PBE0 simulations were carried out to double check our PBE results. 

\emph{Absence of CO$_2$(aq)---}
The simulations just described led us to question the assumption of CO$_2$(aq) 
being the major carbon species in supercritical aqueous fluids in the deep Earth.
We further increased the pressure to $\sim$11 GPa and the temperature to 1000 K, i.e. at pressures similar to those at the bottom of the Earth’s upper mantle, 
to investigate whether our conclusions on the absence of CO$_2$(aq) in water did hold at more extreme conditions. 
The density of water at this P-T condition is 1.57 g/cm$^3$ \cite{pan2013dielectric}.
The equation of state of water in this regime has been investigated in our previous study using the 
PBE functional \cite{pan2013dielectric}. 

Fig. \ref{mole-percent}(B)  shows the mole percentages of dissolved carbon species as functions of simulation time 
obtained in first-principles MD simulations with the PBE and PBE0 functionals. 
We found 41\% CO$_3^{2-}$, 58\% HCO$_3^{-}$ and 1\% H$_2$CO$_3$(aq) when using PBE; 
PBE0 simulations gave 52\% CO$_3^{2-}$ and 48\% HCO$_3^{-}$, as shown in Table. \ref{mol-tab}.  
Similar to our finding at 0.2 GPa and 823 K, PBE yields a higher concentration of HCO$_3^{-}$ than PBE0, 
but the results of the two simulations are closer than at lower P and T. 
Most importantly, less than 1\% CO$_2$(aq) was found in both PBE and PBE0 simulations.

Due to the lack of experimental data, 
many studies at the upper mantle conditions used geochemical models based on extrapolations of available data.
Using extrapolated thermodynamic data, Caciagli and Manning considered CO$_{2}$(aq) as the dominant dissolved carbon species  
in  calcite aqueous solutions up to 1.6 GPa and 1023 K \cite{caciagli2003solubility}.
Based on geochemical models at lower P and T, Manning et al. inferred that bicarbonate ions 
would become less favored relative to CO$_2$(aq) under the mantle conditions \cite{manning2013chemistry}.
However, Facq et al. performed Raman studies of CaCO$_3$ solutions up to 7 GPa and 673 K,
and found that at P $>$ 4 GPa, CO$_3^{2-}$ 
is the dominant carbon species \cite{facq2014situ}.  
Furthermore, based on theoretical models, 
Facq et al. speculated that at higher temperature CO$_2$(aq) may become a dominant species, 
depending on the pressure, though no exact temperature or pressure were reported \cite{facq2014situ}.
In Refs. \cite{caciagli2003solubility, facq2014situ} using solutions at equilibrium with CaCO$_3$ crystals, 
the concentrations of dissolved carbon are lower than that studied in this work; 
it is therefore challenging to detect experimentally the amount of CO$_2$(aq) 
and definitely rule out the existence of molecular CO$_2$. 
Our first-principles MD simulations clearly showed that up to 11 GPa and 1000 K, 
few CO$_3^{2-}$ ions are converted to CO$_2$(aq).

In order to further verify whether CO$_2$(aq) may be stable in water under extreme conditions, 
we directly dissolved CO$_2$ into water at the same P-T condition of the simulation previously described 
\footnote{the carbonate ion was replaced by one CO$_2$ molecule and one water molecule in the same simulation box at $\sim$11 GPa and 1000 K.}, 
and found again that only few percent of CO$_2$(aq) remained after the solution was equilibrated; bicarbonate ions accounted 
for approximately 80\% of dissolved carbon; this result was obtained using both xc functionals as shown in Fig. \ref{mole-percent}(C).

Interestingly, we also found that H$_2$CO$_3$(aq)  was present in the solution with a mole percent  between 11\% and 23\%,
depending on the functionals; note that this concentration is about  20 times larger than that of of CO$_2$(aq) as shown in Table. \ref{mol-tab}.
The formation of H$_2$CO$_3$(aq) via CO$_2$(aq) hydration may follow two pathways \cite{stirling2010h2co3}: 
1) CO$_{2}$(aq) directly reacts with a water molecule to generate H$_2$CO$_3$ (aq); 
2) an intermediate HCO$_3^{-}$ is formed prior to H$_2$CO$_3$(aq) .
At 11 GPa and 1000 K, in both PBE and PBE0 simulations,
we found that H$_2$CO$_3$(aq) forms after HCO$_3^-$ is formed;
however frequent proton hopping events were detected between H$_2$CO$_3$(aq) and water. 
The lifetime of H$_2$CO$_3$(aq) is on the subpicosecond scale. 
The existence of H$_2$CO$_3$ was also found in a 55.5 m CO$_2$ water solution at $\sim$32 GPa and 2000 K 
using first-principles MD with the GGA functional, 
but no concentration was given due to only a 10 ps simulation at the P-T condition \cite{maillet2009ab}.

In aqueous solutions at ambient conditions, H$_2$CO$_3$(aq) exists only in a very low concentration and  
rapidly reverts to CO$_2$(aq) and water \cite{manning2013chemistry}:
 
\begin{equation}
   \mathrm{H_2CO_3(aq) \rightarrow CO_{2}(aq) + H_2O}.
 \label{H2CO3}
\end{equation}
Because it is difficult to detect H$_2$CO$_3$(aq) at ambient conditions, current geochemical models 
can not distinguish  between H$_2$CO$_3$(aq) and  CO$_2$(aq), and 
H$_2$CO$_3$(aq) and  CO$_2$(aq) are modeled as the same species \cite{manning2013chemistry}.
Our findings, however,
point out that 
this conventional treatment may ignore 
the important presence of H$_2$CO$_3$(aq) as a neutral carbon species in the deep Earth.

At ambient conditions, the dissociation reaction of CO$_2$(aq) in pure water: 
\begin{equation}
  \mathrm{CO_{2}(aq) + 2H_2O \rightarrow HCO_3^- + H_3O^+}
\label{carbonic}
\end{equation}
 is rather slow, with an activation free energy of 21.7 kcal/mol (0.94 eV)\cite{wang2009comprehensive, stirling2010h2co3}; 
more than 99\% of CO$_2$ molecules remain intact when the concentration of CO$_2$(aq) is above 0.1 m. 
However, when the hydroxide ion (OH$^-$) is present, 
the activation free energy of the dissociation of CO$_2$(aq) is reduced to 12.0 kcal/mol (0.52 eV). 
As a result, the dissociation reaction is accelerated dramatically;
 it is about 10$^7$ faster than that in pure water \cite{wang2009comprehensive}.
At HP-HT conditions, water molecules are more easily dissociated, producing hydroxide ions, 
which then react with hydrated CO$_2$(aq) thus enhancing its dissociation. 
The enhanced bending motion of C-O bonds at high P-T also facilitates the change of carbon hybridization from $sp$ to $sp^2$.  
In our simulations, the exact concentrations of charged ions depends on the xc functionals used, 
but both the GGA and hybrid functionals give the same result on CO$_2$(aq) concentration: the latter is negligible 
in  aqueous solutions  at 11 GPa and 1000 K.
Hence we conclude that our calculations provide strong and robust evidence of the absence of CO$_2$(aq) in water at this P-T condition.

\begin{figure}
\centering
\vspace{5mm}
\includegraphics[width=0.5 \textwidth]{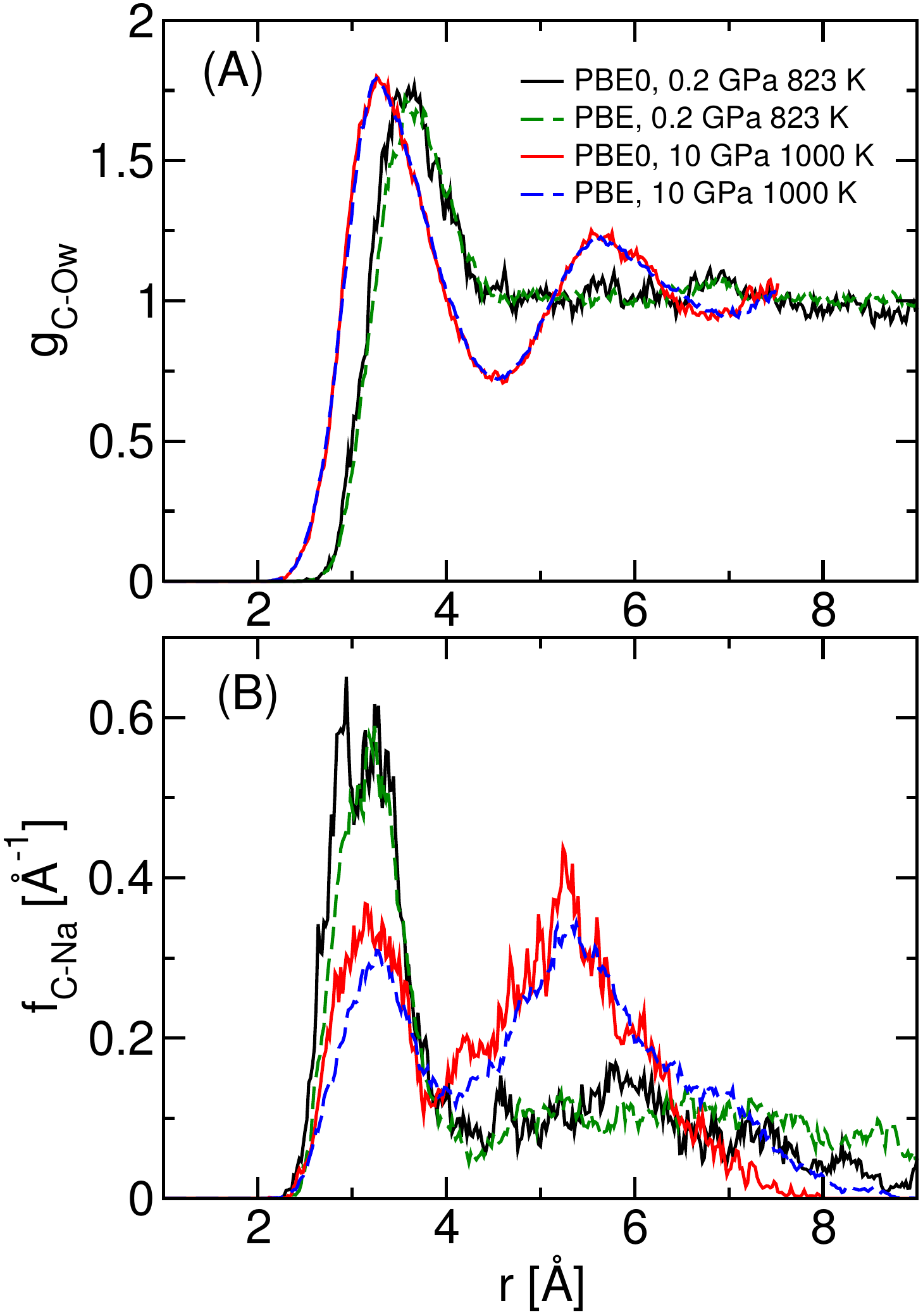}
\caption{(A) Radial distribution functions (RDF) of carbon atoms (C) vs. oxygen atoms of water (O$_w$).
(B) Probability distributions of the distances between carbon atoms and sodium ions.
The MD simulations on the 0.9 m Na$_2$CO$_3$ solutions were conducted at 0.2 GPa, 823 K and 11 GPa, 1000 K. 
Two xc functionals were compared: PBE and PBE0.  
}
\label{rdf}
\end{figure}

\emph{Ion pairing---}
Another important question on the aqueous solutions studied here concerns the
 interactions between (bi)carbonate ions and metal ions.   
In Fig. \ref{rdf}, we plotted the radial distribution functions (RDF) and distributions of the C-Na distances 
obtained from our MD trajectories using the PBE and PBE0 functionals.     
We first discuss the results given by the two functionals, and we then extract robust features of both simulations.

We examined the solution structures of (bi)carbonate ions in the solutions.
In Fig. \ref{rdf}(A), 
the C-Ow (where Ow denotes an O belonging to water molecules) RDFs obtained by PBE and PBE0 are very similar at 0.2 GPa, 823 K or 11 GPa, 1000 K. 
At 11 GPa and 1000 K, the first local minimum of the C-Ow RDFs is at $\sim$4.54 \AA, 
corresponding to the separation of the first and second solvation shells of (bi)carbonate ions. 
Although such a minimum is not easy to define at 0.2 GPa and 823 K, 
since the first peak of the C-Ow RDFs  also appears to vanish at $\sim$4.54 \AA, 
we used the same radial cutoff of 4.54 \AA\ for the first solvation shells of (bi)carbonate ions at two P-T conditions,
in order to have a consistent comparison. 

The ion pairing strength in an aqueous solution greatly depends on the static dielectric constant of water, $\epsilon_0$: the electrostatic interaction between ions in water is $\frac{q_1q_1}{\epsilon_0 r^2}$,
where $q_1$ and $q_2$ are the electrical charges of the ions, and $r$ is the distance between them. 
We note that PBE appears to overestimate  $\epsilon_0$ at ambient conditions, 
while PBE0 yields a result in better agreement with experiment \cite{schoenherr2014dielectric};
hence at the PBE level of theory, the overestimate of electrostatic screening leads to a weaker ion pairing, 
as shown by lower PBE peaks at $\sim$ 3.2 \AA\ in Fig. \ref{rdf}(B).
The interesting thing is that under upper mantle pressure conditions, 
the discrepancy between $\epsilon_0$ evaluated within PBE and the experimental value is rather small \cite{pan2013dielectric},
and we expect PBE to yield more realistic ion associations at those conditions than at ambient conditions.
 
At 0.2 GPa and 823 K, both functionals show marked peaks in the distribution of the C-Na distance at $\sim$3.2 \AA,
indicating strong ion pairing between Na$^+$ and CO$_3^{2-}$ or HCO$_3^{-}$. 
The first peak of the C-Ow RDF is at $\sim$3.5 \AA,
so the Na$^+$-CO$_3^{2-}$/HCO$_3^{-}$ pairs tend to reside inside the first solvation shell of (bi)carbonate ions.
With increasing P and T to 11 GPa and 1000 K respectively, the main peak of the distribution function of the 
C-Na distance decreases in intensity, 
indicating that ion pairing becomes weaker in denser water. 

Table \ref{ion-pairing} shows the mole percents of the ion pairs in the first solvation shells of (bi)carbonate ions.
As shown in Table \ref{mol-tab} and \ref{ion-pairing},
we found that more than 99\% of the CO$_3^{2-}$ anions were accompanied by
one or two Na$^+$ cations in their first solvation shells at 0.2 GPa and 823 K,
while at 11 GPa and 1000 K, less than 63\% of the CO$_3^{2-}$ ions 
had counterions in the first solvation shells. 
This may be attributed to the change of $\epsilon_0$: 
our previous study showed that with increasing pressure, $\epsilon_0$ increases along an isotherm \cite{pan2013dielectric}, 
and thus the dielectric screening is enhanced, and
the electrostatic attraction between Na$^+$ and CO$_3^{2-}$ or HCO$_3^{-}$ is weakened. 
The large dielectric screening also enhances the autoionization of water, 
producing a higher concentration of OH$^-$ ions, whose presence accelerates the dissociation of CO$_2$.

The pairing of  carbonate ions with alkali cations at 0.2 GPa and 823 K 
was also reported in a one-molal K$_2$CO$_3$ solution by using geochemical models \cite{frantz1998raman},
but no spectroscopic evidence was available because of the detection limit \cite{frantz1998raman}.
Recently, Schmidt reported the Raman spectra, which were interpreted as showing Na$^+$-CO$_3^{2-}$ pairing, 
for a 1.6 m Na$_2$CO$_3$ solution interfaced with quartz at T above 673 K.
In the CaCO$_3$-water-NaCl solutions, Facq et al. interpreted the measured Raman data 
by assuming the existence of the NaHCO$_3^\circ$ complex \cite{facq2016carbon}. 
In our study, despite the different ion pairing strengths given by the two functionals, 
we found that ion pairing between Na$^+$ and CO$_3^{2-}$ or HCO$_3^{-}$ indeed occurs at 0.2 GPa and 823 K,
with its strength decreasing at 11 GPa and 1000 K.

\emph{Kinetics of reactions involving CO$_3^{2-}$ and HCO$_3^-$---}
Bicarbonate and carbonate ions are the major species in the solutions studied here, with 
frequent proton transfer over ps time scales. 
Their chemical balance in the Na$_2$CO$_3$ solution  is: 
\begin{equation}
 \mathrm{ CO_3^{2-} + H_2O \rightleftharpoons HCO_3^- + OH^- }.
\label{p-hopping}
\end{equation}
The kinetics of the reaction is key to understanding the chemistry of 
oxidized carbon in supercritical water. 

\begin{figure}
\centering
\vspace{5mm}
\includegraphics[width=0.9 \textwidth]{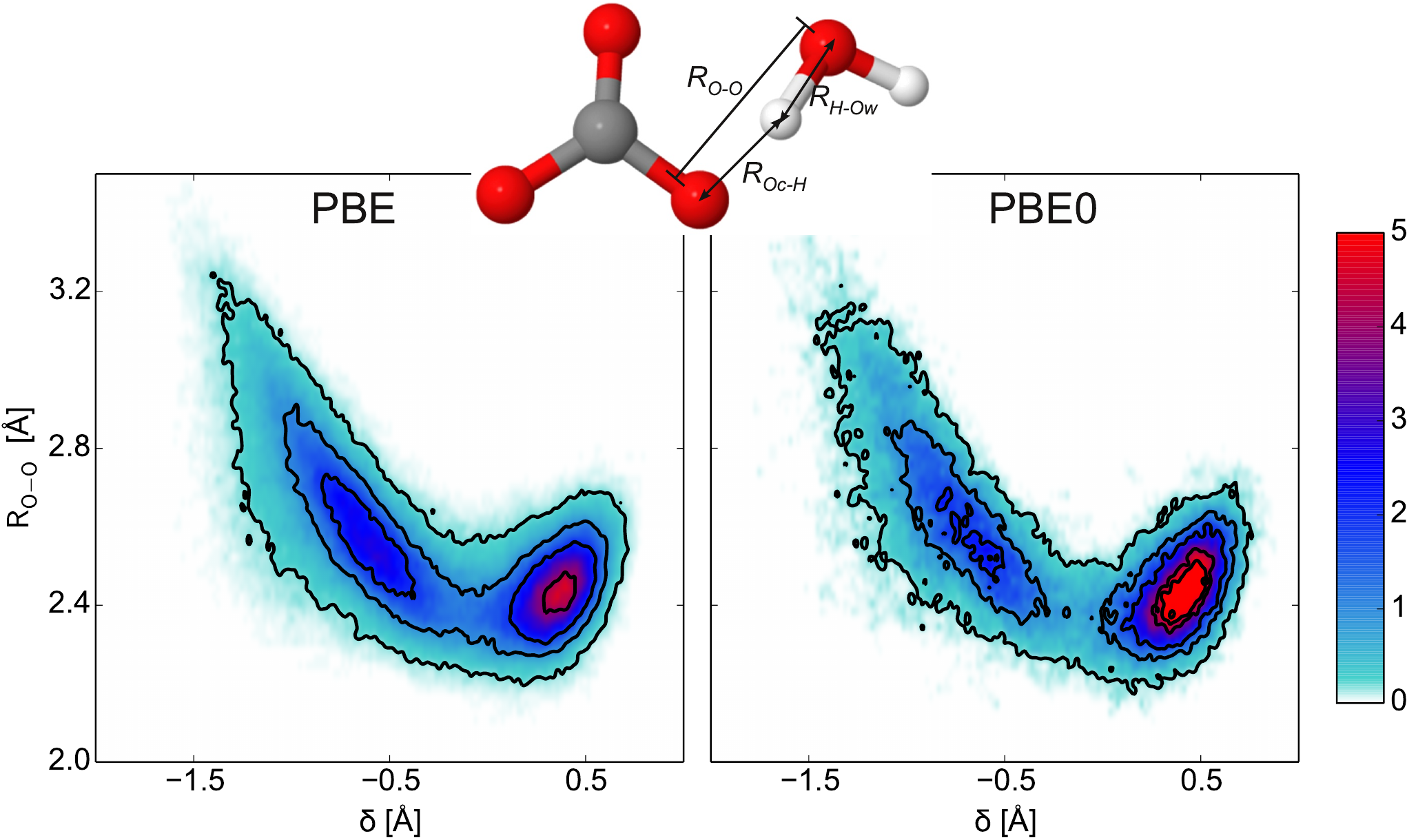}
\caption{Probability distributions of positions of protons hopping between CO$_3^{2-}$ 
and H$_2$O in the Na$_2$CO$_3$ solution at 11 GPa and 1000 K. The unit is \AA$^{-2}$.\ 
The reaction coordinate $R_{O\textnormal{-}O}$ is the distance
between the two neighboring oxygen atoms, O$_c$ and O$_w$, in carbonate ions and water molecules respectively, and
$\delta$ is the proton displacement $R_{Oc\textnormal{-}H}-R_{H\textnormal{-}Ow}$.
Two xc functionals were compared: PBE and PBE0.  
 }
\label{barrier}
\end{figure}

Fig. \ref{barrier} shows the distribution of protons hopping 
between a carbonate ion and its nearest water molecule in the Na$_2$CO$_3$ solution at 11 GPa and 1000 K.
From Fig. \ref{barrier}, we calculated the effective free energy barrier 
of proton transfer to the right in the reaction (\ref{p-hopping}) as
$\Delta F=-k_B T \ln \frac{P_{b}}{P_r}$, where $P_{b}$ and $P_r$ are the probabilities of being on 
the top of the barrier and in the reactant state, respectively, and $k_B$ is
the Boltzmann constant. At the PBE0 level of theory, 
the free energy barrier is estimated to be 4.6 kcal/mol (0.20 eV), while at the PBE level, it decreases to $\sim$ 2.8 kcal/mol (0.12 eV).
According to the Arrhenius equation, the rate constant is $Ae^{-E_a/k_B T}$,
where $E_a$ is the activation energy. 
If we assume PBE and PBE0 yield a similar prefactor $A$, and approximate the activation energy by the free energy barrier, 
the rate constant of the forward reaction (\ref{p-hopping}) using PBE is about three times as large as that found using PBE0.
Such a difference may be explained by the difference in O-H bond strengths obtained using the two functionals.
We computed the vibrational density of states of the solution at 11 GPa and 1000 K in Fig. S1, 
and found that the stretching frequency of O-H bonds given by PBE is about 150 cm$^{-1}$ lower than that obtained with PBE0.
As a result, at the PBE level of theory the O-H bond is weaker and easier to break than at the PBE0 level, 
and the proton transfer rate is enhanced
\footnote{It has been reported that at ambient conditions the vibrational frequencies of the OH stretching mode are also
underestimated using PBE compared with experimental data, whereas PBE0 yields results much closer to experiment \cite{CuiIR2011}. 
The underestimate decreases with increasing temperature  \cite{CuiIR2011}.
Those findings suggest that it is necessary to use hybrid functionals to get accurate reaction rates in aqueous solutions,
especially at ambient conditions.
At extreme conditions, the error of PBE becomes smaller, 
which is again consistent with the results of our previous studies \cite{pan2013dielectric, pan2014refractive}. }.
 
At T = 1000 K, the thermal energy $k_BT$ is 2.0 kcal/mol (0.086 eV), comparable to the free energy barriers given by PBE and PBE0. 
Because of the low free energy barrier and enhanced thermal effects, 
both functionals yield a rate constant of the forward reaction (\ref{p-hopping}) about  six orders of magnitude faster than that at ambient conditions
\footnote{The rate constant of the forward reaction (\ref{p-hopping}) 
 is 3.06$\times$10$^5$ s$^{-1}$ in sea water at ambient conditions \cite{schulz2006determination}.}, 
indicating highly active ionic interactions in our solutions at 11 GPa and 1000 K.

It is worth noting that Fig. \ref{barrier} shows only part of the reaction (\ref{p-hopping}); 
the proton transfer is not only limited to the distance between a carbonate ion and a neighboring water molecule,
but it may involve multiple water molecules connected by a hydrogen bond (HB) wire, via a Grotthuss mechanism \cite{marx2006proton, geissler2001autoionization}.  
In our solutions, protons move back and forth along  HB wires over short time scales ($<$0.1 ps), leading to short-lived (bi)carbonate ions.
However, if the thermal fluctuation breaks HB wires, protons can not hop back and then bicarbonate ions persist for few picoseconds.
The forming and breaking of the HB wires are critical to the lifetime of carbon species in  the calculations reported here.

In Raman experiments, the symmetric stretching mode of C-O/C-OH at $\sim$ 1000 cm$^{-1}$ is used to detect 
carbonate or bicarbonate ions. The vibrational period of this mode is 33 fs. 
For the carbon species, whose lifetime is comparable to 33 fs or even shorter, 
the Raman peaks are broadened considerably, and may be difficult to identify experimentally.

In our MD simulations the forces acting on nuclei
were computed quantum mechanically, but protons were still treated as classical particles. 
If quantum nuclear effects are taken into account, we expect the proton transfer rate to be enhanced, 
though at such high temperatures quantum effects may not be as important as at ambient conditions \cite{ceriotti2013nuclear}.

Using both PBE and PBE0 functionals, van der Waals  interactions are described inaccurately; 
however, in our study we focused on the breaking and forming of covalent bonds 
and the lack of dispersion forces is not expected to affect our main conclusions. 
In addition we note that results for the equilibrium volume and dielectric constant of ice obtained with PBE and van der Waals functionals agree much better at high pressure (e.g. for ice VIII) than at low pressure (e.g. for ice Ih and Ic) \cite{PhysRevLett.108.105502}, indicating that the lack of dispersion forces may affect results for water and ice more at ambient conditions than under pressure.

Widely used geochemical models based on the Born function, e.g., the Helgeson–Kirkham–Flowers (HKF) model as implemented in the SUPCRT92 \cite{johnson1992supcrt92}, 
consider water as a continuum medium with no molecular structure  taken into account, when calculating thermodynamic
properties of electrolytes \cite{Anderson_book}.
In our study we found that the microscopic properties of water at the molecular scale, e.g.,  proton transfer mechanisms, 
play a substantial role in the chemical reactions of oxidized carbon, particularly at the high P-T conditions studied here.

\section*{Conclusion}
In conclusion, 
we carried out first-principles molecular dynamics simulations to investigate dissolved carbonate and CO$_2$ in water 
at high pressure and high temperature, 
up to the conditions of the Earth's upper mantle.
Contrary to popular geochemical models assuming that CO$_2$(aq) 
is the major carbon species present in water,
we found that most of the dissolved carbon at 11 GPa and 1000 K 
is in the form of solvated CO$_3^{2-}$ and HCO$_3^-$ anions.
The two anions exchange protons with water on a picosecond time scale. 
The proton transfer along hydrogen bond wires driven by thermal fluctuations 
is responsible for the fast dynamics observed in these solutions. 
Under such extreme conditions, even H$_2$CO$_3$(aq) is more abundant than CO$_2$(aq), 
a dramatic difference with respect to ambient conditions. 
While it is well known that the autoionization of water is greatly enhanced at extreme conditions,
e.g., at the P-T conditions of the Earth's mantle, and that the pH of neutral water may be well below 7\cite{liebscher2010aqueous}, 
it has been long unclear how the notable changes of water properties under pressure affect the solvation of oxidized carbon.
we showed, using atomistic ab initio simulations, that the coordination number of carbon changes in water
at the Earth's upper mantle conditions.  
We also found that ion pairing between Na$^+$ 
and CO$_3^{2-}$ or HCO$_3^-$ is greatly affected by P-T conditions, 
decreasing with pressure at 800 $\sim$ 1000 K.
Our results suggest that at  extreme conditions water
transports carbon mostly through highly active ions, not CO$_2$(aq) molecules. 

The results reported here give important insights into deep carbon science. 
Our calculations predicted that the carbon dissolved in water-rich geofluids 
is in ionic forms in the bottom of the Earth's upper mantle, indicating that
CO$_2$ degassing may mostly occur close to the Earth's surface.

 \section*{Methods}
First-principles molecular dynamics (MD) simulations were performed in the Born-Oppenheimer approximation 
using the Qbox code (http://qboxcode.org/) \cite{gygi2008architecture}.  
Two exchange-correlation (xc) functionals were used: the semilocal functional PBE and the hybrid functional PBE0.
In order to improve the efficiency of our calculations using PBE0, 
the recursive subspace bisection method was used with a threshold of 0.02 \cite{gygi2009compact, gygi2012efficient}. 
The Brillouin zone of the supercell was sampled with the $\Gamma$ point only.
We used norm-conserving pseudopotentials \cite{HSC, vanderbilt1985}\footnote{Pseudopotential Table: http://fpmd.ucdavis.edu/potentials}
with a kinetic cutoff of 85 Ry, which was increased to 220 Ry for pressure calculations.  
The MD time step was 0.24 fs. 
Deuterium was used instead of hydrogen,  so as to use a larger time step for computational convenience.  
Note that the density given in the main text was computed for light water. 
The temperature was maintained constant by the Bussi-Donadio-Parrinello thermostat ($\tau=24.2$ fs)\cite{bussi2007canonical}.

To determine whether CO$_2$ or CO$_3^{2-}$ were present in our simulations, 
we searched the atomic trajectories generated in our simulations for the three closest 
O atoms to a given C atom, and ordered these O atoms according to their respective C-O distances.
If the C-O distance of the third closest O atom was larger than that of the second one
by more than 0.4 \AA, the species was considered as CO$_2$, otherwise it was defined as a CO$_3^{2-}$ anion.
The C-O bond length in all sets of our MD runs was found to be 1.3$\pm$0.2 \AA. 
For H atoms around the carbonate ion, 
we sought the nearest O atom to each of them, and if such O atom 
belonged to the carbonate ion, we concluded that a bicarbonate ion, HCO$_3^{-}$, was formed. 
When two H atoms were bonded to CO$_3^{2-}$ simultaneously, the solute then became carbonic acid.  

Using the criteria described above, we determined the type of solute molecules present in our MD simulations at each time step.
Note that at HP-HT conditions, covalent bonds are frequently broken and reformed; 
hence in order to establish how the concentration of the solute varied along 
our MD trajectories, we calculated 
the mole percents of carbon species as functions of simulation time.
The mole percent of the $i$th species at time $t$ was obtained as:
\begin{equation}
x_i(t) = \frac{n_i(t)}{N_w}\times 100\%,
\end{equation} 
where $n_i(t)$ is the number of the snapshots containing the $i$th species 
between the time ($t-\tau_w$) and $t$, 
and $N_w$ is the total number of snapshots in such a time interval. 
Here, $\tau_w$ was set to 1 ps.  

\section*{SUPPLEMENTARY MATERIALS}
fig. S1. The vibrational density of states of the 0.9 m Na$_2$CO$_3$ solution at $\sim$11 GPa and 1000 K.
\\
fig. S2. Probability distributions of positions of protons hopping between CO$_3^{2-}$ 
and H$_2$O in the Na$_2$CO$_3$ solution at 0.2 GPa and 823 K.

\section*{Acknowledgements}
We thank F. Giberti and D. Gygi for many useful discussions and D. A. Sverjensky for a critical reading of the manuscript and many discussions. 
This work was supported by the  Alfred P. Sloan Foundation through the Deep Carbon Observatory (DCO) and by DOE-BES grant No. DE-SC0008938. 
Part of this work was carried out using the DCO Computer Cluster, 
and  resources of the Argonne Leadership Computing Facility (ALCF), 
provided by an award of computer time  by the Innovative and Novel Computational Impact on Theory and Experiment (INCITE) program. 
ALCF is a DOE Office of Science User Facility supported under Contract DE-AC02-06CH11357.
We also acknowledge the University of Chicago Research Computing Center for support of this work.

\section*{Contributions}
D.P. and G.G. designed the research. 
Calculations were performed by D.P.
All authors contributed to the analysis 
and discussion of the data and the writing of the manuscript. 

\section*{Competing Interests} 
The authors declare that they have no competing interests.

\section*{Data and materials availability}
 All data needed to evaluate the conclusions in the paper are present in the paper and/or the Supplementary Materials. Additional data related to this paper may be requested from the authors.

\newpage

\begin{table}
\caption{Carbon species in the 0.9 m Na$_2$CO$_3$ and 0.9 m CO$_2$ solutions in first-principles molecular dynamics simulations. 
The concentrations are shown as mole percents of the total dissolved carbon. 
The mole percents of CO$_3^{2-}$ and HCO$_3^-$
include those of ion pairs formed by them, respectively.
The experimental results (Expt.)  are from the Raman data in Ref. \cite{frantz1998raman} (see text).} 
\vspace{5mm}
\begin{tabular}{|c|cc|c|cccc|}
\hline
\multirow{2}{*}{Solution}&\multirow{2}{*}{P[GPa]}&\multirow{2}{*}{T[K]}&\multirow{2}{*}{Method}&\multicolumn{4}{|c|}{Carbon species [\%]}\\
\cline{5-8}
& & & &CO$_2$&CO$_3^{2-}$&HCO$_3^-$&H$_2$CO$_3$\\
\hline
\multirow{5}{*}{Na$_2$CO$_3$}&\multirow{3}{*}{0.2}&\multirow{3}{*}{823}&PBE&0.6&16.7&82.4&0.1\\
& & &PBE0&0.2&26.8&72.9&0.1\\
& & &Expt.&N/A&32&68&N/A\\
\cline{2-8}
&\multirow{2}{*}{11}&\multirow{2}{*}{1000}&PBE&0.1&40.8&57.9&1.2\\
& & &PBE0&0.1&51.8&47.5&0.6\\
\hline
\multirow{2}{*}{CO$_2$}& \multirow{2}{*}{11}&\multirow{2}{*}{1000}&  PBE &0.6&8.2&79.8&11.4\\
& & & PBE0 &0.0&1.7&75.0&23.3\\
\hline

\end{tabular}
\label{mol-tab}
\end{table}

\begin{table}
\caption{The ion pairing in the first solvation shells of the carbon species 
in the 0.9 m Na$_2$CO$_3$ solutions. The concentrations are shown as mole percents of the \emph{total} dissolved carbon.} 
\vspace{5mm}
\begin{tabular}{|cc|c|cccc|}
\hline
\multirow{2}{*}{P[GPa]}&\multirow{2}{*}{T[K]}&\multirow{2}{*}{Method}&\multicolumn{4}{|c|}{Ion pairing species [\%]}\\
\cline{4-7}

& & &NaCO$_3^-$&Na$_2$CO$_3^\circ$&NaHCO$_3^\circ$&Na$_2$HCO$_3^+$\\
\hline
\multirow{2}{*}{0.2}&\multirow{2}{*}{823}&PBE&3.7&13.0&59.7&7.5\\
& &PBE0&8.4&18.2&40.8&18.2\\
\hline
\multirow{2}{*}{11}&\multirow{2}{*}{1000}&PBE&18.1&5.9&28.9&7.5\\
& &PBE0&22.4&10.0&20.4&12.5\\
\hline
\end{tabular}
\label{ion-pairing}
\end{table}

\end{document}